\documentclass{article}

\usepackage{arxiv}
\usepackage{amsmath}
\usepackage[linesnumbered,ruled,vlined]{algorithm2e}
\usepackage[utf8]{inputenc} 
\usepackage[T1]{fontenc}    
\usepackage{hyperref}       
\usepackage{url}            
\usepackage{booktabs}       
\usepackage{amsfonts}       
\usepackage{nicefrac}       
\usepackage{microtype}      
\usepackage{lipsum}
\usepackage{graphicx}
\graphicspath{ {./images/} }

\title{FAIRTOPIA: Envisioning Multi-Agent Guardianship for Disrupting Unfair AI Pipelines}
\author{
 Athena Vakali \\
  School of Informatics\\
  Aristotle University of Thessaloniki\\
  Thessaloniki, Greece \\
  \texttt{avakali@csd.auth.gr} \\
   \And
Ilias Dimitriadis \\
  School of Informatics\\
  Aristotle University of Thessaloniki\\
  Thessaloniki, Greece \\
  \texttt{idimitriad@csd.auth.gr} \\
}
\makeatletter
\renewcommand{\@maketitle}{
  \vbox{%
    \hsize\textwidth
    \linewidth\hsize
    \vskip 0.1in
    \@toptitlebar
    \centering
    {\LARGE\sc \@title\par}
    \@bottomtitlebar
    \textsc{}\\
    \vskip 0.1in
    \def\And{%
      \end{tabular}\hfil\linebreak[0]\hfil%
      \begin{tabular}[t]{c}\bf\rule{\z@}{24\p@}\ignorespaces%
    }
    \def\AND{%
      \end{tabular}\hfil\linebreak[4]\hfil%
      \begin{tabular}[t]{c}\bf\rule{\z@}{24\p@}\ignorespaces%
    }
    \begin{tabular}[t]{c}\bf\rule{\z@}{24\p@}\@author\end{tabular}%
  }
}
\makeatother

\begin{document}
\maketitle
\begin{abstract}
AI models have become active decision makers, often acting without human supervision. The rapid advancement of AI technology has already caused harmful incidents that have hurt individuals and societies and AI unfairness in  heavily criticized. It is urgent to disrupt AI pipelines which largely neglect human principles and focus on computational biases exploration at the data (pre), model(in), and deployment (post) processing stages. We claim that by exploiting the advances of agents technology, we will introduce cautious, prompt, and ongoing fairness watch schemes, under realistic, systematic, and human-centric fairness expectations. We envision agents as fairness guardians, since agents learn from their environment, adapt to new information, and solve complex problems by interacting with external tools and other systems. To set the proper fairness guardrails in the overall AI pipeline, we introduce a fairness-by-design approach which embeds multi-role agents in an end-to-end (human to AI) synergetic scheme. Our position is that we may design adaptive and realistic AI fairness frameworks, and we introduce a generalized algorithm which can be customized to the requirements and goals of each AI decision making scenario. Our proposed, so called FAIRTOPIA framework, is structured over a three-layered architecture, which encapsulates the AI pipeline inside an agentic guardian and a knowledge-based, self-refining layered scheme. Based on our proposition, we enact fairness watch in all of the AI pipeline stages, under robust multi-agent workflows, which will inspire new fairness research hypothesis, heuristics, and methods grounded in human-centric, systematic, interdisciplinary, socio-technical principles. 
\end{abstract}


\section{Introduction}
\begin{figure}[ht!]
\centering
\includegraphics[width=0.8\textwidth, height=3.5cm]{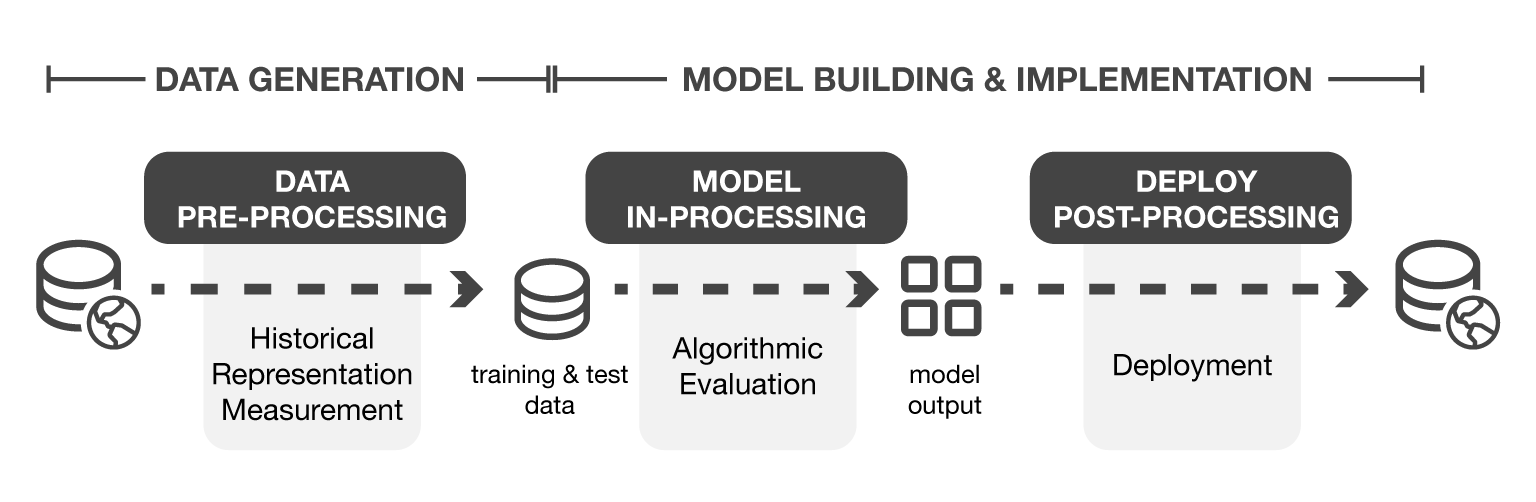}
\caption{Indicative biases detected at the AI pipeline stages.}
\label{fig:AI-pipeline}
\end{figure}

The enduring importance of fairness in shaping human behavior and social structures
\cite{fairness_evolution_Brosnan14} is strongly challenged in our emerging human-and-AI-symbiotic era.
Fairness is essential to safeguard adoption of trusted AI systems which will ensure that \textit{AI can make the right and fair decisions}. Such an endeavor is too ambitious, since AI comes with diverse requirements, risks, and expectations that vary in different domains and ecosystems.
For example, AI fairness should be employed rather differently when we target for healthcare equitable diagnoses and treatment recommendations, loan approvals based on impartial financial factors, customer service chatbots provide consistent and unbiased support, AI recommends products, etc. \cite{sutton2018digitized, AV19}. Thus, AI fairness is a challenging and nuanced delicate matter which demands \textit{deep understanding of the context} of each addressed decision making scenario, and no "one solution fits all" approach may tackle it \cite{vereschak2024trust}. Achieving fairness in AI-driven decision making has become the “holy grail” of an overwhelmed body of prominent research with numerous families of methods (supervised or unsupervised learning, statistics, graph based, clustering, etc.), multiplicity of metrics, along with a continuous thread of surveys \cite{AAG24, CH24, F24, HCZ24}. The most popular approaches explore unfair or unjust outcomes over AI pipelines,   engineered to presume specific bias types over the sequential stages of data collection (pre-processing), methods adaptation (in-processing), and deployment (post-processing) \cite{MMS21,CH24}, as depicted in Figure \ref{fig:AI-pipeline}. Such a vast amount of research propositions lacks a systematic and integral formulation and is criticized for the ad hoc and controversial solutions produced.

The current solutions for fair AI are mostly reactive, post-hoc, and overly statistical, fail to address the systemic, evolving, and interactive nature of biases in real-world scenarios. Deploying fair AI is by default targeting systems without bias a practically too difficult and even infeasible goal. Research evidence and a vivid debate about the \textbf{(im)possibility of Fairness} \cite{MCB23, BBD23, CIN23, FSV21} has set theoretical questions which demand for feasible and realistic fairness goals. Since AI driven decision making has to balance even contradicting fairness properties (eg. individuals versus groups limitations), it is no longer accepted that an AI solution is more fair if it is less biased. Contradictions in real world scenarios are amplified since humans are also biased, and thus keeping a balance among human and AI requirements becomes too complicated. Up to now, the major limitations in AI driven decision making is due to : lack of contextual understanding and subjective tradeoffs (e.g., group vs. individual fairness); neglect of human cognitive biases reflected into datasets, labels, model design, and deployment choices; and over-reliance on AI (such as in supervised learning) which is tested over benchmark datasets, often away from the actual domain-specific fairness requirements and goals.

\noindent\textit{[Position Statements:]}
We argue that current AI fairness methodologies are fundamentally unprepared for the emerging changes of our ecosystems which rapidly become heavily AI-dependent. Such a universal adoption of AI introduces new unprecedented risks of unfairness due to the misalignment among the evolution of technology and the inclusion of human principles.  
We claim that we must explore novel ways to safeguard fairness beyond the strict AI pipeline stages, and monitor AI harms in dynamic environments of varying fairness risk levels. To this end, we propose a paradigm shift which is urgently needed to transform the AI pipeline from its current sequential biases detection norm (Fig. \ref{fig:AI-pipeline}). Our so called \texttt{FAIRTOPIA} socio-technical framework aims to guard fairness and detect bias interrelations along with human oversight when and if needed, under realistic -and non utopian- fairness expectations. We will address the above emerging critical fairness challenges by exploiting the emerging agents technology which is characterized by its disruptive autonomous, adaptive, and dymamic decision making capacity. Thus, our \texttt{FAIRTOPIA} proposition addresses the next critical Research Questions (\textit{[RQs]}).\\
\noindent[\textit{RQ1: How to turn human and AI fairness evidence} into a knowledge base which will safeguard fairness in all AI pipeline stages ?] $\rightarrow$ We propose a new design and deployment plan to build a fairness knowledge base in machine-processable forms, based on rich research and information about the causes and effects of human and AI bias in the real world.
\\
\noindent [\textit{RQ2: Which mechanisms can monitor fairness leakage and risks} in dynamic human and AI coaligned decision making ?] $\rightarrow$ \texttt{FAIRTOPIA} introduces new interactive and iterative mechanisms based on actionable agents that will adapt knowledge elements and tools into contextualized fairness guardrails to carefully monitor fairness in the overall AI pipeline. 
\\ 
\noindent [\textit{RQ3: Is it feasible to reform AI pipelines into  fairness by design workflows able to adapt to complex and varying human and AI symbiotic scenarios?}]$\rightarrow$ \texttt{FAIRTOPIA} offers a new architecture for implementing multi-agent end-to-end fair pipelines and ensure fairness cautious AI workflows.\\
We respond to the above \textit{[RQs]}, with novel disruptive propositions detailed in Sections 3, by building on the rationale of existing knowledge and technology readiness (summarized in Section 2), and by an adaptive and flexible framework design which is described in Section 4, along with our proposition's critique and advocacy summarized in Section 5. 

\section{Fairness in the midst of emerging AI technologies} 
As discussed above, realistic fairness expectations should be highly contextualized and well positioned with respect to each underlying AI decision making scenario. Such an endeavour must be driven by credible research and technological evidence. We claim that there is credible \textbf{fairness hidden wisdom}, given the historic fairness quest in the human evolution, which has revealed several types of bias in cognitive and psychology science. Earlier research work has produced hundreds of cognitive bias types that interfere with how we process information, think critically, and perceive reality cautiously or unconsciously \cite{PH23, IS14}. Over than half a century work, recognized by Nobel prize \cite{TK74}, has offered very rich scientific evidence and has identified many cognitive biases as a systematic and common error in human judgement. Recognizing such human (biased) blind spots is highly complex \cite{PH23, FSV21} and even more complex is to formulate and measure these human biases. Since AI is built and governed by humans, it is reasonable (even unconsciously) to transfer and reflect these biases into each process of the AI pipeline, causing unpredictable harm \cite{D23, GHF22}. The human cognitive heuristics and their reflection on AI computational biases is surprisingly not studied in a complete manner. Current computational AI fairness methods largely neglect human and societal (systemic) biases, which are much more complex and constitute the root of the problem, found at the bottom of a recently introduced so-called 'biases iceberg' \cite{SVG22}. The proposition of taxonomies of AI harms, risks, and incidents, such has the Algorithmic, and Automation Harms \cite{ABG24}, highlights the need for a human-centric cautious computational fairness monitoring. \\
In parallel, we witness a \textbf{remarkable burst of AI technologies}, such as Large Language Models (LLMs) which brought conversational AI into the lives of millions in a matter of months, and emerging Generative AI (GenAI) technologies which rapidly catalyze the broader AI ecosystem \cite{IS27}. LLM-inspired discoveries have refedined fundamental constructs, such as Knowledge Graphs (KGs), but have also introduced novel technologies, such as Retrieval-Augmented Generation (RAGs), which dynamically retrieve relevant information from large repositories, unlike standalone LLMs that rely solely on pre-trained knowledge \cite{GXG23, J23, RY24}. Combinations of all such AI-relevant technologies (LLMs, RAGs, KGs, etc) has taken multiple forms: RAGs and KGs enrich LLMs, with up-to-date and contextually relevant information, improving accuracy and informativeness, and mitigate significant LLMs limitations (spurious biases, hallucinations, non-factual reasoning, etc.), while KGs rich factual information, enhance LLMs inference and interpretability \cite{YGZ24, J23, PXN23}; 
RAGs – LLMs – KGs fusion has introduced Graph RAGs, by using LLMs to build graph-based indices forming KGs from source documents, and then pre-generate community summaries for all groups of closely-related entities \cite{ETC24}; The integration of LLMs and KGs can take various forms : KGs can be useful during the pre-training and inference phases of LLMs, or for enhancing the understanding of knowledge learned by LLMs. Additionally, KGs can be augmented by using LLMs for various KG tasks, such as KG embeddings (KGEs to represent KG elements into a continuous vector space), completion, construction, graph-to-text generation, and question answering. 
KGs taxonomies have played an enhanced role in the introduction of Comprehensible Artificial Intelligence (CAI), as the parent concept of Explainable Artificial Intelligence (XAI) and Interpretable Machine Learning (IML), offering advanced data model hierarchies display abilities to improve both human and machine readability \cite{SWS23}. 

In parallel to LLMs and GenAI technologies which have taken the world by storm, a yet more transformative technology, the so called “Agentic AI”, is showcasing its potential \cite{SAB23, QZG24}. 'AI agents' increasingly serve as automated personal assistants since they deploy AI modules/systems that can autonomously plan and execute complex online tasks, with limited human oversight \cite{CSM23}. The opportunities for disrupting AI fairness by this new technology are tremendous while its momentum and potential are clear and promising \cite{K24, QZG24, VP24}. We argue that bias monitoring in AI must embark on our meta-GenAI era, well beyond typical LLMs, since unprecedented new technologies emerge with very promising bias-awareness and fairness safeguarding capabilities. Addressing bias and fairness in such a rapid evolution of GenAI research demands novel and out-of-the-box thinking, given concerns already expressed about LLMs' ability to achieve fairness \cite{CZC23, DGD23, BL24}. 

\section{Agents of fairness : harvesting knowledge and raising guardrails}
Our proposition is based on credible evidence to address fairness by scalable and adaptive human and AI alignment. Bias and fairness \textbf{constructs of knowledge} along with \textbf{guardian technologies} will introduce novel schemes for self-revised fairness-aware and trustworthy AI pipelines. We argue that genuine fairness in AI decision making settings will be achieved only by leveraging interdisciplinary insights to design and deliver actual evidence and knowledge which is now fragmented and published by different scientific communities (social, computer, psychology, policy making). Thus, we revisit the \textit{Fairness research hypothesis in AI}, and we propose a knowledge-based to encapsulate all AI computational processes, into a dynamically adjusted bias monitoring and fairness guardrailing scheme. Next, we summarize the knowledge generation process and the principles of our agentic approach, while the proposed framework's architecture is presented in Section 4. 

\subsection{Building a Fairness Knowledge Base}
\begin{figure}[ht!]
\centering
\includegraphics[width=0.8\textwidth, height=4.5cm]{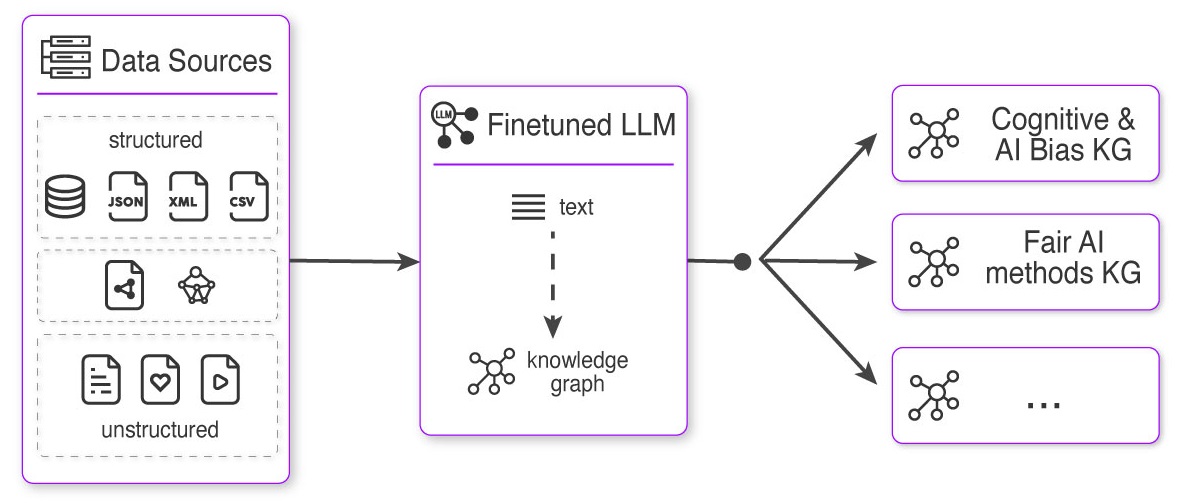}
\caption{Generation of multiple Knowledge Graphs based on heterogeneous sources}
\label{fig:RP1-KGs}
\end{figure}
As discussed in Section 2, the abundance of currently fragmented information about cognitive and AI bias and fairness, may offer valuable knowledge and reveal human bias persistence into the inner AI pipeline (pre, in, post) stages. To uncover human to AI bias and fairness cause and effects, we now have the structures and the technology to build a systematic knowledge base, a kind of a \textit{fairness thesaurus}. 
We introduce a systematic wisdom base to connect the dots of existing intense inter-disciplinary research about bias and fairness, by conceptualization of bias and its role on fairness schemes in machine processable forms such as metric spaces, Knowledge Graphs (KGs), and embedding based repositories. 
To systematically correlate currently fragmented cognitive versus AI research evidence \cite{SVG22}, we introduce a flexible design and an implementation plan for an extensible \textbf{Fairness knowledge "warehouse"} of information sources which will harvest knowledge coming from multiple information sources. Indicatively, such sources include: (a) numerous earlier fairness published research work; (b) abundance of actual datasets, models, metrics, outcomes in computational fairness quest paradigms in research and in practice; (c) cognitive and AI bias and harmful incidents taxonomies along with bias critical harms cases; (d) APIs, Bias and Fairness documentation repositories; and (e) any other online credible and authoritative sources.
As depicted in Figure \ref{fig:RP1-KGs}, the proposed 'knowledge warehouse' will form a credible base that will act as an input to cutting-edge advanced GenAI technologies, such as LLMs and KGs. As indicated in Section 2, we have multiple options to integrate and advance current out-of-the-self methods for automating KGs construction. We are now able to fine tune LLMs which will assist in multi (cognitive bias – data – models – metrics – etc) KGs generation. This process will enable systematic vectorization and indexing of multiple types of cognitive and computational bias in machine-processable forms, by embedding models \cite{JLC23, CFM24}. 
By advancing recent research for KG completion and harnessing LLMs' inference proficiency and KG structure-aware reasoning, we will fine-tune LLMs with KG Alignment models (e.g. via neighborhood partitioning or Generative Subgraph Encoding \cite{ZCG24, DAZ24}.
We aim for a systematic theoretically grounded approach for authoritative KGs, which offers fertile ground for human-machine readable solutions \cite{SWS23, ZWL23}, which may be exploited in different AI decision-making scenarios.
Depending on each AI scenario, researchers can use hundreds of the latest methods in KG construction \cite{ZWL23}, including advanced techniques in knowledge acquisition, KG completion, knowledge fusion, knowledge reasoning, and high-quality KG embeddings \cite{JLC23, PXN23, CFM24}. Furthermore, given that addressing AI fairness has not exploited RAGs technology adequately, a few frameworks like FairRAG, have been evaluated solely on image data \cite{SZC24} and have not yet proven their capacities. We propose to thoroughly explore and propose the synergistic integration of LLMs, RAGs, and KGs (outlined in Section 2). In summary, our main proposition is to design and generate an extensible and theoretically grounded bias and fairness knowledge base, which will reveal currently hidden and largely overlooked patterns. By introducing fairness cautious LLMs fine-tuning schemes, multiple KGs and fairness embeddings and vectorized datasets may be generated and adapted, in each of the AI  fairness task's scenario and requirements. Representing human and AI biases as KGs will advance cutting edge LLMs to further construct and optimize mathematical formulation and an algorithmic approach, as discussed in the next subsection.

\subsection{Agents as AI fairness guardrail generators}
To align the evidence from our proposed "Knowledge warehouse" with realistic fairness goals, we propose an Agentic scheme, based on the emerging agents generation wave, which redefines our human-AI symbiotic reality (as discussed in Section 2). 

\begin{algorithm}[htbp]
\SetAlgoLined
\LinesNumbered
\KwData{Task $T_{input} :[FG_T:$Fairness Goals; $max(trials)$; $KGs$; $M_{\textit{set}}$];}
\KwResult{$Guardrail_{set}$ for each AI pipeline stage; $Best_{outcome}$ optimized $T_{\textit{plans}}$ sets.}
\textit{// initialization phase} \\ 
$F_{pla}$: planner agent for pre-processing; \; 
$F_{act}$: action agent for in-processing; \; 
$F_{\textit{opt}}$: optimizer agent for post-processing;\; 
$T_{plans} \gets T_{\textit{input}}$\;  
$t \gets 0$\;
\textit{// operational phase} \\
\While{$F_{opt}$ fails on $FG_T$ or $t<max(trials)$}{
  $F_{pla} \gets [FG_T; T_{plans}]$ and produce-fairness($Guardrail_{set}$)\;
  $F_{act} \gets [Guardrail_{set};T_{plans}]$ and recommend($Best_{outcome}$)\;
  \uIf{$Best_{outcome}$ fails on $FG_T$}{
    self-critic($Best_{outcome} \gets [KGs,M_{set}]$)\;
  }
  $F_{opt} \gets Best_{outcome}$ and reflect-optimize($T_{plans}$)\;
  $t \gets t+1$\;
}
\caption{Fairness Agentic Alignment algorithm}
\label{alg:align}
\end{algorithm}

We introduce a generalized AI fairness alignment Algorithm \ref{alg:align}, which gets multiple inputs to grasp all  contextualized information required for each AI Task $T_{input}$. Next, we highlight the main phases of our algorithm and showcase its generalization and adaptation capacity.

\noindent[\textit{{Input}}]: each task $T{input}$ comes with its own contextualized information which includes specifications elements, in the form of sets of rules on datasets access; restrictions lists posed by an organization; individuals or groups privacy levels; etc. This contextualized information must be specified by each implementor (organization, developers, decision bodies, etc.) of Task $T_{input}$, since each scenario comes with its distinct fairness goals $FG_T$. Setting realistic fairness expectations is vital, since varying AI decision-making scenarios come with their own particular fairness risks. Such risks are already categorized under various policy making regulations, such as the EU AI Act, which has classified AI risk levels as "unacceptable", "high", "limited" and "minimal" \cite{IS25}. 
Our algorithm is open for its employment over GenAI and LLMs technology settings, since the above input will be formulated into query sets about data schemes, methods' goals, fairness risks, case studies fairness risk requirement lists, descriptors, along with the KGs and vectorized data from our proposed "Fairness warehouse" and any other structured knowledge (e.g. bias taxonomies, AI harms hierarchies) and $M_{sets}$ of open code, tools, and software libraries.\\
\noindent[\textit{{A modular Algorithmic approach }}]: our algorithm's principle is that each AI task $T_{input}$ may be iteratively refined with optimized $T_{plans}$, under the synergy of three distinct agents which will safeguard each of the corresponding three (pre, in, post) AI pipeline stages. Thus, we ensure proper pre-processing planning with the $F_{Pla}$ agent, in-processing action with the $F_{act}$ agent, and proceed to post-processing optimization with the $F_{opt}$ agent. Under a systematic sequence of operations, our three core $F_{pla}$, $F_{act}$, and $F_{opt}$ proposed agents, will continuously assess the fairness goals $FG_{T}$, specified per each Task $T_{input}$, along with producing agentic-spotted guardrails as $Guardrail_{set}$, which will enable cautious and continuous bias and fairness watch in all of the AI pipeline stages. Such a multi-agentic interaction iteratively self-reflects and refines fairness choices and balances human oversight and AI automation with an adaptive and respectful to $FG_T$ approach, as indicated in the Algorithm \ref{alg:align}. The proposed guardrails generation is vital to avoid sustaining "black-box" algorithms which lead to unfair and mistrusted outcomes. Our proposed generalized algorithm will resolve current AI fairness bottlenecks such as neglecting valuable knowledge about cognitive and AI biases correlations, associating actual patterns and impact of the many biases and spotting contradicting fairness risks and expectations. 

\begin{figure}[ht!]
\centering
\includegraphics[width=0.6\textwidth, height=6.0cm]{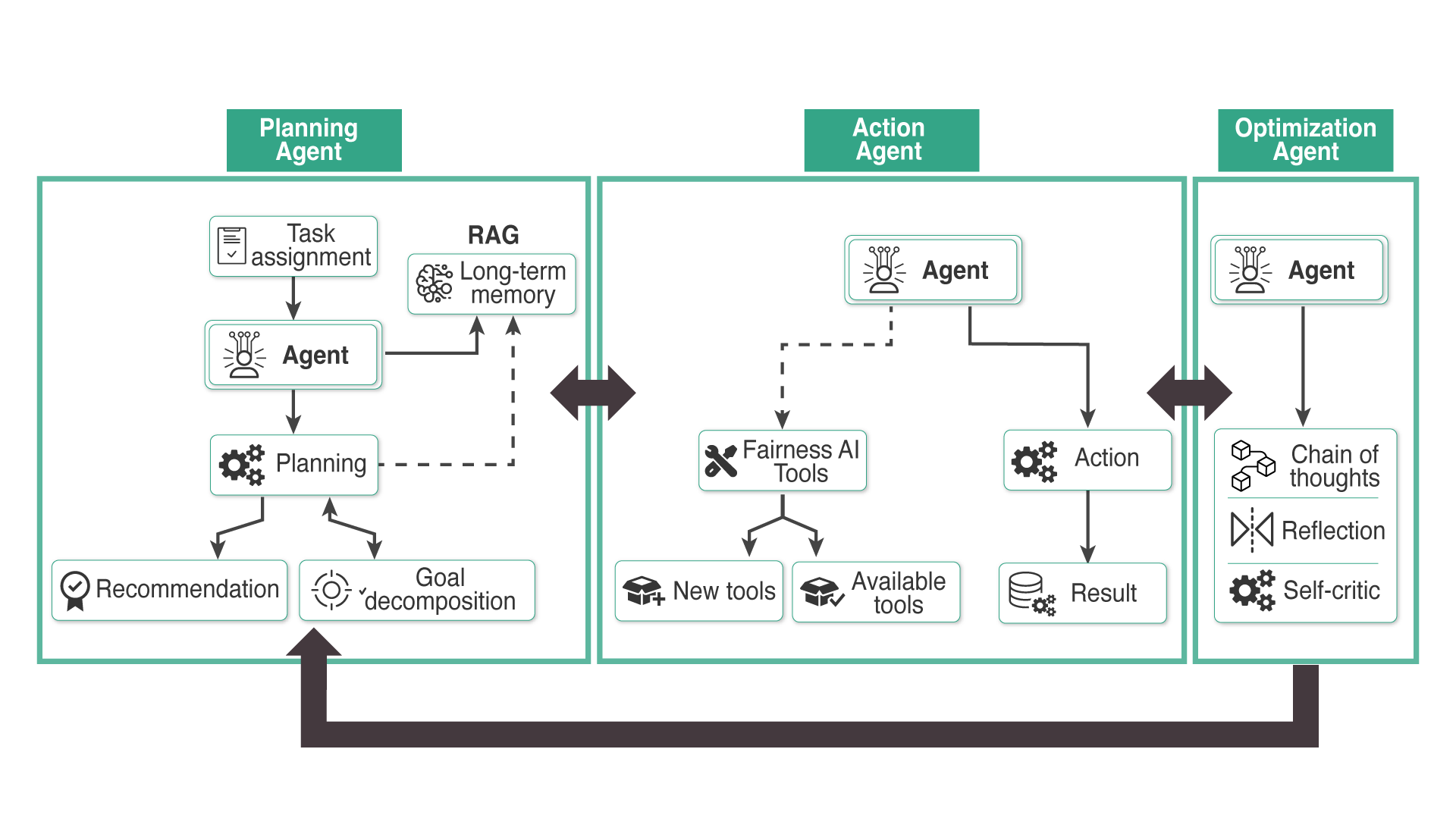}
\caption{Agents for fair AI planning, deployment and optimization.}
\label{fig:AI-agents}
\end{figure}

Our proposed approach inspires cautious and realistic \textit{fair AI algorithmic steps}, driven by the proposed distinct agents which are designed to autonomously pursue complex goals and workflows with limited direct human supervision. 
The proposed AI fairness-aware agents will operate based on the cutting edge agentic techniques, as outlined in Figure \ref{fig:AI-agents}.

The three agents have distinct roles, and their synergy will build an agentic layer, designed to autonomously pursue complex goals with limited direct human supervision. To achieve their goals, agents will embed and extend state of the art dynamic and self-reflective processes 
\cite{SCG24, CSM23, SAB23}. Given the complexities and challenges involved in achieving realistic fair AI-driven decisions with a synergetic human-AI manner, the operation of the proposed agents is expected to introduce new views and move GenAI forward beyond simple request and response tactics. Up to now, only  recent and limited work has addressed the issue of fairness challenges and harms in Agentic AI settings \cite{RagaSaki25,Z24}. Our proposed AI bias-aware agents are able to engage in dynamic and self- reflective processes, based on the Algorithm \ref{alg:align} Input specifications and the realistic requirements.
Appropriate planning and recommendations will come up by setting the $F_{pla}$ agent to receive each task's requirements (in the form of $T_{input}$) and exploit the proposed knowledge warehouse (multiple $KG_s$), to deliver guardrail sets ($Guardrail_{set}$) and refined new tasks ($Task_{plans}$ recommendations under continuous decomposition and cross-checking of $FG_{T}$. Advanced research will be conducted to optimize fair AI planning, not only by optimizing $Tasks_{plans}$, but also by assessing new recommendations $Best_{\textit{outcome}}$. By advancing RAGs and the relevant GenAI technology (Section 2) to store and assess varying agentic fairness schemes, this line of research will balance human and AI interactions in designing and implementing adaptive and fine-tuned agents \cite{QZG24, SCG24}. The implementation of agents $F_{act}$ and $F_{opt}$ will ensure the continuous monitoring of fairness at the AI in- and post- processing stages, based on their flexible, extensible, and refined schemes (outlined in Fig. \ref{fig:AI-agents}). In particular, $F_{act}$ has the capacity to exploit all the elements suggested by $F_{pla}$ and allow for multiple methods ($M_{set}$) and AI fairness toolsets under repeated actions of implementing algorithms and supporting the several fairness plans produced. Finally, the agent $F_{opt}$ will implement recently advanced refinement techniques in agent systems, such as the chain of thoughts, reflection, and self-critic \cite{ZYZ25}, \cite{ZTW24, VVM25}, to produce $Best_{\textit{outcome}}$.

\section{FAIRTOPIA: A tri-partite Framework to safeguard AI Fairness}

\begin{figure}[ht!]
\centering
\includegraphics[width=0.9\textwidth, height=7.5cm]{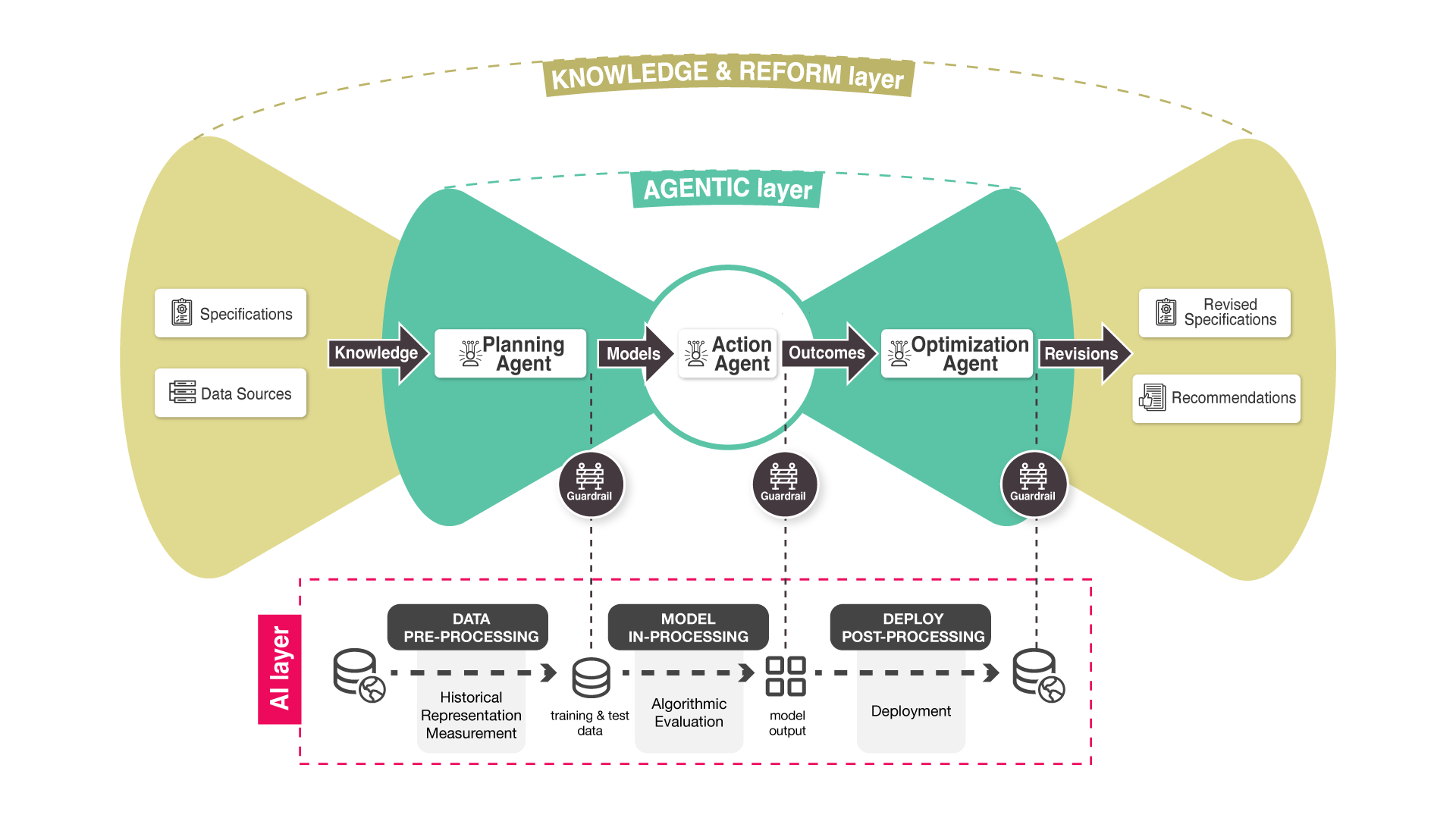}
\caption{Agentic AI fairness framework : architecture overview}
\label{fig:architecture}
\end{figure}

Our proposed approach minimizes fairness risks and sets a solid base for deploying and refining actual fairness in real world AI decision making workflows. We introduce a “fairness by design” architecture, which disrupts the typical AI (pre, in, post) pipeline, and we safeguard fairness leakage end-to-end, building on the credible knowledge base (Section 3) and the agentic refinements algorithm. Given the complexities and challenges involved in current fairness in AI and the emerging synergetic demands of human-AI, our agentic AI alignment goes beyond sequential request and response and aims at a human-AI alignment under iterative and interactive AI processes, to be implemented under the proposed \texttt{FAIRTOPIA} flexible and robust framework. \textbf{FAIRTOPIA} introduces a new sociotechnical framework that unfolds in three layers (depicted in Figure \ref{fig:architecture}). 

\noindent 

This layered architecture embeds the typical AI pipeline (Fig. \ref{fig:AI-pipeline}) as the \textit{AI layer} which is not isolated in its own processing stages, but which interacts with the \textit{Agentic} and the \textit{Knowledge and Reform} layers, which produce the proper fairness quardrails at all stages of the AI pipeline. The role and specifics of each layer are summarized in the following.  
\begin{itemize}
 \item \textbf{AI layer} : includes the typical AI pipeline stages that are no longer sequential but are disrupted by fairness leakage guardrails produced by the other two layers. The AI pipeline is not an ad-hoc fairness testbed at which bias is measured per stage, but it embeds the reasoning, the origins, and the fairness coming from the overall cognitive (human) to computational (AI) inherent interactions. 
  \item \textbf{Agentic layer} : includes all of our three agents ($F{pla}; F_{act};F_{opt}$) implementations, such that they interact with each other but also with iteratively obtain input and produce relevant results. As discussed in Section 3, $F{pla}$ will get as input the KGs which will materialize the general knowledge base (Sect. 3.1) and produce as output the proper fairness models which are mostly tailored for the underlying AI decision making task. The functionality of $F{pla}$ will follow the proposed algorithmic steps (Sect. 3.2) in order to iteratively flag the guardrails generated and refine the models details (data; code; metrics; etc.). Thus, the $F_{act}$ agent will proceed to the execution of methods which have been filtered with respect to their fairness capacity, and will also follow a method refinement repetitive approach in which algorithmic and evaluation fairness guardrails are produced. These guardrails will safeguard the input given to the $F_{opt}$ agent which will assess the actual fairness of deployment, reached under reflective and self-critic rounds, to produce guardrails for the overall AI pipeline fairness. The  guardrails of the overall agentic layer will be communicated and intervene with both the AI and the Knowledge-Reform layers. 
 \item \textbf{Knowledge and Reform layer} : includes an "umbrella'' to cover bias and fairness knowledge harvested up to the reformed agentic recommendations, enabling iterative cycles which materialize the proposed modular multi-agents (planning, action, optimization). Crafting self-refined plans and action rounds is a demanding endeavour since we must repeatedly self-criticize the chosen plans and actions, and sustain a cycle of repeated planning-action-refine loops towards delivering $Best_{\textit{outcome}}$. Novel sophisticated setups will be needed to leverage a continuum of the $\text{produce-fairness}$; $\text{recommend}$; and $\text{produce}$ methods indicated in Algorithm \ref{alg:align} pseudocode.
\end{itemize}
We argue that genuine fairness in AI systems can be achieved only by leveraging interdisciplinary insights, and by recognizing bias pathways from human (cognitive) to the AI (computational) sides. Thus, the proposed \texttt{FAIRTOPIA} architecture overcomes siloed and monolithic AI pipeline implementation restrictions and enables steering and optimizing multiagent pipelines with emphasis on human-centric principles. We envision human and AI hybrid pipelines at which human oversight is carefully embedded in all the proposed layers, inspired by emerging human and AI symbiotic views \cite{PPS24}. Human oversight will be explored by approaches such as the so called Human-in-the-loop (HITL) model or Interactive Machine Learning (IML) methods to ensure that AI does not evoke adverse effects against humans, and to foster compliance with relevant fairness regulations \cite{AC24, GCC24}. These approaches will ensure that humans have an important and interactive role in the various stages of AI, and their role will be carefully placed based on the guardrails produced in the agentic AI layer. Human inclusiveness at all \texttt{FAIRTOPIA} layers will be encouraged on a context-based and interactive manner driven by emerging IML techniques which balance coalition of both experts (e.g. bias, AI experts, data scientists) and non-experts (lay public, citizens groups) when and if needed. Involvement of humans will take place only when the respective guardrails indicate their need to prevent fairness leakage, and the proposed optimized refined plans will adopt robust combination of HITL approaches and collaborative scientific methodologies. Human participation will thus contribute to any of the stages (eg. data collection specifications, data analysis interpretation, bias cautious recommendations) and offer insightful elements in drawing genuine conclusions. The importance of HITL in the resolution of LLM problems, such as hallucinations, is already highlighted by Agentic AI interested communities, who point out that HITL balance is important and further research will be required to detect an optimal HITL strategy. In our Agentic layer, we will ensure that HITL involvement will not reduce automation and that HITL will be integrated at our Agentic layer to produce substantial outcomes and refined guardrails on which to rely. In \texttt{FAIRTOPIA}, humans may be involved, for instance, as bias annotators, as algorithm and outcome validators, as critics to balance accuracy and reliability of algorithmic plans, as iterative feedback estimators, etc. Involving individuals or groups of humans will demand a proper choice of relevant skills, and we propose an incrementally updated methodology \cite{BCH22}, based on which humans in fairness watch loops will perform tasks in which they are more efficient than AI. Such an approach will ensure bidirectional feedback loops (human - AI) under the principles of quality and ethical assurance \cite{MHA23}.

\texttt{FAIRTOPIA} framework allows the implementation of different and diverse validation scenarios under proper AI fairness tasks specifications, and is extensible for multiple deployment and experimentation rounds. The datasets to be used, the metadata and features prioritized and the domain-specific constraints will drive the overall multi-layer interaction. Multiple varying fairness-risk scenarios may be planned  and each of them is expected to demand proper fine-tuning and adaptive tasks preparation and monitoring, under the optimization capacity of our multi-agentic AI workflows. Interactions among layers will enable fairness quadrails to be based on credible authoritative KGs and to receive refined recommendations and proceed to iterative actions. Bias-aware and fairness-aligned data schemes will receive each case study  specific data profiling, constraints, and sources selection and will turn original data sources into bias-aware fairness aligned data and information for proper training-testing datasets, under an ongoing  action-refine agentic dialogue. 
\texttt{FAIRTOPIA} architecture follows a \textbf{bowtie-like design}, which is particularly useful in risk analysis as it visually represents (in a bowtie shape), both the causes (left side) and consequences (right side) of a central event (the core in the middle). As depicted in Figure \ref{fig:architecture}, our core represents the actual AI model operational processes with all the corresponding $F_{\textit{act}}$ agent's functionality, while our proposed knowledge base and actions of $F_{\textit{pla}}$ agent is set as the causal (left) side of the bowtie, and the refinement steps of the $F_{\textit{opt}}$ agent are shown as the resulting effects at the right side of the bowtie structure.
Our approach promises algorithmic bias resilience since its guardrails adapt models requirements, as well as algorithmic and deployment strategies to eventually deliver the most tailored solutions per AI underlying scenario. Furthermore, the agentic monitored bias and fairness risks allow for human-centric an open extension to explainable AI assessments for providing trustworthy and comprehensive summarization of metrics for fairness and performance best outcomes. Such an adaptive workflow will integrate bias and fair AI reflections, driven by the planning-action-refine agents to advance powerful meta Gen-AI technologies enabling human comprehensive inclusion and efficient oversight. In summary, our proposition disrupts the typical AI pipeline to overcome its current monolithic bias mitigation and fairness safeguarding, with an ambitious Agentic AI framework which will bridge and unite currently fragmented research evidence from cognitive and AI scientific communities.

\section{Counterarguments and Discussion}
\texttt{FAIRTOPIA} proposes fairness as a continuously negotiated, system-integrated principle rather than an external audit criterion. Its key innovations include : Graph-based representations for structured knowledge mapping and transparency; interactive agents that simulate fairness deliberation and contextual adaptation; a flexible architecture which embeds HITL and IML loops integrated at all stages with clear oversight touchpoints.

Surely, the proposed solution requires extensive and deep interdisciplinary research, as formulating and automating fairness entities is a rather complex and challenging issue. Research is needed for new \textit{axioms and theorems} that will govern projections and inter-dependencies of human and AI biases, with the goal of resulting in optimized quadrails and refined plans. Numerous AI computational solutions, from now on, must go deeper into understanding the inner fair algorithmic logic. Moreover, the very recent work by pioneers in AI refers to catastrophic risks posed by the so called "Superintelligent Agents'', which demand for new  safer \textit{Scientist AI} solutions \cite{bengio25}.
\texttt{FAIRTOPIA} recognizes the core critique elements and aims to contribute in such a line of research and it advocates with its indicative next research orientation suggestions and insights.  \\ 
\textbf{Critique 1:} Agentic AI is still experimental and existing AI pipelines will not adapt easily to the proposed multilayer approach.\\
\textit{Advovacy:} Agentic systems already deploy in real-world settings (e.g., LLM agents, autonomous assistants), and their autonomy and feedback loops will soon prove existing fairness AI strategies inadequate. The use of agentic AI is rapidly expanding, with predictions that by 2028, 33\% of enterprise software applications will incorporate it \cite{agentic-stat}. Improvements in all technologies discussed in Section 2 (Large Language Models, memory capabilities, reinforcement learning, and tool integration) are further enabling agentic AI factors which will need to achieve greater autonomy and safeguard fairness complex risks.

\textbf{Critique 2:} Modeling and correlating cognitive and AI biases is infeasible.\\
\textit{Advocacy:} Decades of social science and cognitive research have already begun to provide robust taxonomies which can be encoded in computational forms such as the use of KGs and embeddings (suggested in Section 3). We also suggest that viewing cognitive and AI biases over metric spaces will ignite further fruitful research in the area. We may define a Cognitive space $CS$ as a metric space noted as $CS=(B,d_b)$, where $B$: bias types and $d_b$: distances between them (estimated by \cite{TK74} suggestions) and assume that distance $d_b$ captures closeness of each bias type to an underlying AI Task $T_{input}$. Similarly, if an AI space $AS$ is a metric space defined as $AS=(\hat{B},\hat{d})$, a bias reflection function will estimate the properties of $r:B\rightarrow\hat{B}$ and its ability to produce bias entities $\hat{b}=r(b)$ from a bias $b\in\ CS$, and its corresponding distances $\hat{d}$. Bias in social and cognitive sciences (formulated under our KGs) along with the abundance of fair AI solutions systematic constructs, will enable human and AI bias types similarities detection over both \textit{CS},\textit{AS}) spaces, and harvest their reflections, hierarchies, patterns, and dynamics. 

\textbf{Critique 3:} The field lacks maturity for standardization.\\
\textit{Advovacy:} Interdisciplinary methodologies, HITL designs, and community-driven taxonomies will be needed to allow standardization of fairness constructs in adaptive contexts. However, some early implementations have been formulated with agentic AI in domains such as customer support automation, HR operations, and IT support. Furthermore, the exploration of fairness in high-risk domains, such as in health, offer already specific 'Fairness by Design'' guidelines which are open for adoption and further systematic standardization for designing, developing, evaluating and deploying human-centered health monitoring sensing technologies \cite{AV16}.

Thus, \textit{FAIRTOPIA} opens a research and extensive multi-domain dialogue for fairness novel quest at the lab (synthetic) and real-world (in the wild) settings. We envision a highly sustainable, compute- and data-intensive architecture to enable domain agnostic and domain specific configurations towards robust and realistic benchmarking. In summary, we urge the community to : shift toward agent-aware, socio-technical fairness paradigms; develop and share standardized fairness relevant knowledge graphs and evaluation benchmarks; promote interdisciplinary research across social, cognitive psychology, data and AI sciences; design adaptive human-centric frameworks with feedback-driven fairness evolution.



\begin{thebibliography}{10}

\bibitem{fairness_evolution_Brosnan14}
S.~F. Brosnan and F.~B. De~Waal.
\newblock Evolution of responses to (un) fairness.
\newblock {\em Science}, 346(6207), 2014.

\bibitem{sutton2018digitized}
Andrew Sutton, Reza Samavi, Thomas~E Doyle, and David Koff.
\newblock Digitized trust in human-in-the-loop health research.
\newblock In {\em 2018 16th Annual Conference on Privacy, Security and Trust (PST)}, pages 1--10. IEEE, 2018.

\bibitem{AV19}
Sofia Yfantidou, Pavlos Sermpezis, Athena Vakali, and Ricardo Baeza-Yates.
\newblock Uncovering bias in personal informatics.
\newblock {\em Proceedings of the ACM on Interactive, Mobile, Wearable and Ubiquitous Technologies}, 7(3):1--30, 2023.

\bibitem{vereschak2024trust}
Oleksandra Vereschak, Fatemeh Alizadeh, Gilles Bailly, and Baptiste Caramiaux.
\newblock Trust in ai-assisted decision making: Perspectives from those behind the system and those for whom the decision is made.
\newblock In {\em Proceedings of the CHI Conference on Human Factors in Computing Systems}, pages 1--14, 2024.

\bibitem{AAG24}
Tosin Adewumi, Lama Alkhaled, Namrata Gurung, Goya van Boven, and Irene Pagliai.
\newblock Fairness and bias in multimodal ai: A survey.
\newblock {\em arXiv preprint arXiv:2406.19097}, 2024.

\bibitem{CH24}
S.~Caton and C.~Haas.
\newblock Fairness in machine learning: A survey.
\newblock {\em ACM Computing Surveys}, 2024.

\bibitem{F24}
Emilio Ferrara.
\newblock Fairness and bias in artificial intelligence: A brief survey of sources, impacts, and mitigation strategies.
\newblock {\em Sci}, 6(1):3, 2023.

\bibitem{HCZ24}
Max Hort, Zhenpeng Chen, Jie~M Zhang, Mark Harman, and Federica Sarro.
\newblock Bias mitigation for machine learning classifiers: A comprehensive survey.
\newblock {\em ACM Journal on Responsible Computing}, 1(2):1--52, 2024.

\bibitem{MMS21}
Ninareh Mehrabi, Fred Morstatter, Nripsuta Saxena, Kristina Lerman, and Aram Galstyan.
\newblock A survey on bias and fairness in machine learning.
\newblock {\em ACM computing surveys (CSUR)}, 54(6):1--35, 2021.

\bibitem{MCB23}
Suvodeep Majumder, Joymallya Chakraborty, Gina~R Bai, Kathryn~T Stolee, and Tim Menzies.
\newblock Fair enough: Searching for sufficient measures of fairness.
\newblock {\em ACM Transactions on Software Engineering and Methodology}, 32(6):1--22, 2023.

\bibitem{BBD23}
Andrew Bell, Lucius Bynum, Nazarii Drushchak, Tetiana Zakharchenko, Lucas Rosenblatt, and Julia Stoyanovich.
\newblock The possibility of fairness: Revisiting the impossibility theorem in practice.
\newblock In {\em Proceedings of the 2023 ACM Conference on Fairness, Accountability, and Transparency}, pages 400--422, 2023.

\bibitem{CIN23}
Daniele Regoli, Alessandro Castelnovo, Nicole Inverardi, Gabriele Nanino, and Ilaria Penco.
\newblock Fair enough? a map of the current limitations of the requirements to have fair algorithms.
\newblock {\em arXiv preprint arXiv:2311.12435}, 2023.

\bibitem{FSV21}
Sorelle~A Friedler, Carlos Scheidegger, and Suresh Venkatasubramanian.
\newblock The (im) possibility of fairness: Different value systems require different mechanisms for fair decision making.
\newblock {\em Communications of the ACM}, 64(4):136--143, 2021.

\bibitem{PH23}
Emily Pronin and Lori Hazel.
\newblock Humans’ bias blind spot and its societal significance.
\newblock {\em Current Directions in Psychological Science}, 32(5):402--409, 2023.

\bibitem{IS14}
World~Economic Forum.
\newblock Cognitive biases that are warping your perception of reality, 2021.
\newblock World Economic Forum report.

\bibitem{TK74}
Amos Tversky and Daniel Kahneman.
\newblock Judgment under uncertainty: Heuristics and biases.
\newblock 1990.

\bibitem{D23}
Virginia Dignum.
\newblock Responsible artificial intelligence---from principles to practice: A keynote at thewebconf 2022.
\newblock In {\em ACM SIGIR Forum}, volume~56, pages 1--6. ACM New York, NY, USA, 2023.

\bibitem{GHF22}
Benjamin Van~Giffen, Dennis Herhausen, and Tobias Fahse.
\newblock Overcoming the pitfalls and perils of algorithms: A classification of machine learning biases and mitigation methods.
\newblock {\em Journal of Business Research}, 144(6):93--106, 2022.

\bibitem{SVG22}
Reva Schwartz, Reva Schwartz, Apostol Vassilev, Kristen Greene, Lori Perine, Andrew Burt, and Patrick Hall.
\newblock {\em Towards a standard for identifying and managing bias in artificial intelligence}, volume~3.
\newblock US Department of Commerce, National Institute of Standards and Technology~…, 2022.

\bibitem{ABG24}
Gavin Abercrombie, Djalel Benbouzid, Paolo Giudici, Delaram Golpayegani, Julio Hernandez, Pierre Noro, Harshvardhan Pandit, Eva Paraschou, Charlie Pownall, Jyoti Prajapati, et~al.
\newblock A collaborative, human-centred taxonomy of ai, algorithmic, and automation harms.
\newblock {\em arXiv preprint arXiv:2407.01294}, 2024.

\bibitem{IS27}
{European Research Council}.
\newblock Erc science journalism initiative, 2024.
\newblock \url{https://erc.europa.eu/sites/default/files/content/pages/pdf/ERC-Science-Journalism-Initiative.pdf}.

\bibitem{GXG23}
Yunfan Gao, Yun Xiong, Xinyu Gao, Kangxiang Jia, Jinliu Pan, Yuxi Bi, Yixin Dai, Jiawei Sun, Haofen Wang, and Haofen Wang.
\newblock Retrieval-augmented generation for large language models: A survey.
\newblock {\em arXiv preprint arXiv:2312.10997}, 2:1, 2023.

\bibitem{J23}
Cheonsu Jeong.
\newblock Generative ai service implementation using llm application architecture: based on rag model and langchain framework.
\newblock {\em Journal of Intelligence and Information Systems}, 29(4):129--164, 2023.

\bibitem{RY24}
Keshav Rangan and Yiqiao Yin.
\newblock A fine-tuning enhanced rag system with quantized influence measure as ai judge.
\newblock {\em Scientific Reports}, 14(1):27446, 2024.

\bibitem{YGZ24}
Bo~Yuan, Shenhao Gui, Qingquan Zhang, Ziqi Wang, Junyi Wen, Bifei Mao, Jialin Liu, and Xin Yao.
\newblock Fairerml: An extensible platform for analysing, visualising, and mitigating biases in machine learning [application notes].
\newblock {\em IEEE Computational Intelligence Magazine}, 19(2):129--141, 2024.

\bibitem{PXN23}
Ciyuan Peng, Feng Xia, Mehdi Naseriparsa, and Francesco Osborne.
\newblock Knowledge graphs: Opportunities and challenges.
\newblock {\em Artificial Intelligence Review}, 56(11):13071--13102, 2023.

\bibitem{ETC24}
Darren Edge, Ha~Trinh, Newman Cheng, Joshua Bradley, Alex Chao, Apurva Mody, Steven Truitt, Dasha Metropolitansky, Robert~Osazuwa Ness, and Jonathan Larson.
\newblock From local to global: A graph rag approach to query-focused summarization.
\newblock {\em arXiv preprint arXiv:2404.16130}, 2024.

\bibitem{SWS23}
Simon Schramm, Christoph Wehner, and Ute Schmid.
\newblock Comprehensible artificial intelligence on knowledge graphs: A survey.
\newblock {\em Journal of Web Semantics}, 79:100806, 2023.

\bibitem{SAB23}
Yonadav Shavit, Sandhini Agarwal, Miles Brundage, Steven Adler, Cullen O’Keefe, Rosie Campbell, Teddy Lee, Pamela Mishkin, Tyna Eloundou, Alan Hickey, et~al.
\newblock Practices for governing agentic ai systems.
\newblock {\em Research Paper, OpenAI}, 2023.

\bibitem{QZG24}
Yuxiao Qu, Tianjun Zhang, Naman Garg, and Aviral Kumar.
\newblock Recursive introspection: Teaching language model agents how to self-improve, 2024.

\bibitem{CSM23}
Alan Chan, Rebecca Salganik, Alva Markelius, Chris Pang, Nitarshan Rajkumar, Dmitrii Krasheninnikov, Lauro Langosco, Zhonghao He, Yawen Duan, Micah Carroll, Michelle Lin, Alex Mayhew, Katherine Collins, Maryam Molamohammadi, John Burden, Wanru Zhao, Shalaleh Rismani, Konstantinos Voudouris, Umang Bhatt, Adrian Weller, David Krueger, and Tegan Maharaj.
\newblock Harms from increasingly agentic algorithmic systems.
\newblock In {\em 2023 ACM Conference on Fairness Accountability and Transparency}, page 651–666. ACM, June 2023.

\bibitem{K24}
Andrei Kucharavy.
\newblock Fundamental limitations of generative llms.
\newblock In {\em Large Language Models in Cybersecurity: Threats, Exposure and Mitigation}, pages 55--64. Springer Nature Switzerland Cham, 2024.

\bibitem{VP24}
Bart~S Vanneste and Phanish Puranam.
\newblock Artificial intelligence, trust, and perceptions of agency.
\newblock {\em Academy of Management Review}, (ja):amr--2022, 2024.

\bibitem{CZC23}
Shiyao Cui, Zhenyu Zhang, Yilong Chen, Wenyuan Zhang, Tianyun Liu, Siqi Wang, and Tingwen Liu.
\newblock Fft: Towards harmlessness evaluation and analysis for llms with factuality, fairness, toxicity, 2024.

\bibitem{DGD23}
Dwivedi~S Dwivedi~S, Ghosh~S.
\newblock Gender fairness in llms using prompt engineering and in-context learning.
\newblock {\em Rupkatha Journal on Interdisciplinary Studies in Humanities}, 2023.

\bibitem{BL24}
Marion Bartl and Susan Leavy.
\newblock From 'showgirls' to 'performers': Fine-tuning with gender-inclusive language for bias reduction in llms, 2024.

\bibitem{JLC23}
Zhifeng Jia, Haoyang Li, and Lei Chen.
\newblock Air: Adaptive incremental embedding updating for dynamic knowledge graphs.
\newblock In {\em International Conference on Database Systems for Advanced Applications}, pages 606--621. Springer, 2023.

\bibitem{CFM24}
Jiahang Cao, Jinyuan Fang, Zaiqiao Meng, and Shangsong Liang.
\newblock Knowledge graph embedding: A survey from the perspective of representation spaces.
\newblock {\em ACM Computing Surveys}, 56(6):1--42, 2024.

\bibitem{ZCG24}
Yichi Zhang, Zhuo Chen, Lingbing Guo, Yajing Xu, Wen Zhang, and Huajun Chen.
\newblock Making large language models perform better in knowledge graph completion.
\newblock In {\em Proceedings of the 32nd ACM International Conference on Multimedia}, pages 233--242, 2024.

\bibitem{DAZ24}
Stefan Dernbach, Khushbu Agarwal, Alejandro Zuniga, Michael Henry, and Sutanay Choudhury.
\newblock Glam: Fine-tuning large language models for domain knowledge graph alignment via neighborhood partitioning and generative subgraph encoding.
\newblock In {\em Proceedings of the AAAI Symposium Series}, volume~3, pages 82--89, 2024.

\bibitem{ZWL23}
Lingfeng Zhong, Jia Wu, Qian Li, Hao Peng, and Xindong Wu.
\newblock A comprehensive survey on automatic knowledge graph construction.
\newblock {\em ACM Computing Surveys}, 56(4):1--62, 2023.

\bibitem{SZC24}
Robik Shrestha, Yang Zou, Qiuyu Chen, Zhiheng Li, Yusheng Xie, and Siqi Deng.
\newblock Fairrag: Fair human generation via fair retrieval augmentation.
\newblock In {\em Proceedings of the IEEE/CVF Conference on Computer Vision and Pattern Recognition}, pages 11996--12005, 2024.

\bibitem{IS25}
European Commission.
\newblock The eu artificial intelligence act “up-to-date developments and analyses of the eu ai act, 2024.

\bibitem{SCG24}
Noah Shinn, Federico Cassano, Ashwin Gopinath, Karthik Narasimhan, and Shunyu Yao.
\newblock Reflexion: Language agents with verbal reinforcement learning.
\newblock {\em Advances in Neural Information Processing Systems}, 36:8634--8652, 2023.

\bibitem{RagaSaki25}
Rajesh Ranjan, Shailja Gupta, and Surya~Narayan Singh.
\newblock Fairness in multi-agent ai: A unified framework for ethical and equitable autonomous systems.
\newblock {\em arXiv preprint arXiv:2502.07254}, 2025.

\bibitem{Z24}
W.~Zhang.
\newblock Ai fairness in practice: Paradigm, challenges, and prospects.
\newblock {\em Ai Magazine}, 45(3):386--395, 2024.

\bibitem{ZYZ25}
Zhuosheng Zhang, Yao Yao, Aston Zhang, Xiangru Tang, Xinbei Ma, Zhiwei He, Yiming Wang, Mark Gerstein, Rui Wang, Gongshen Liu, et~al.
\newblock Igniting language intelligence: The hitchhiker’s guide from chain-of-thought reasoning to language agents.
\newblock {\em ACM Computing Surveys}, 57(8):1--39, 2025.

\bibitem{ZTW24}
Wenqi Zhang, Ke~Tang, Hai Wu, Mengna Wang, Yongliang Shen, Guiyang Hou, Zeqi Tan, Peng Li, Yueting Zhuang, and Weiming Lu.
\newblock Agent-pro: Learning to evolve via policy-level reflection and optimization.
\newblock {\em arXiv preprint arXiv:2402.17574}, 2024.

\bibitem{VVM25}
Simon Vanneste, Astrid Vanneste, Kevin Mets, Tom De~Schepper, Ali Anwar, Siegfried Mercelis, and Peter Hellinckx.
\newblock Learning to communicate using a communication critic and counterfactual reasoning.
\newblock {\em Neural Computing and Applications}, pages 1--18, 2025.

\bibitem{PPS24}
Clara Punzi, Roberto Pellungrini, Mattia Setzu, Fosca Giannotti, and Dino Pedreschi.
\newblock Ai, meet human: Learning paradigms for hybrid decision making systems.
\newblock {\em arXiv preprint arXiv:2402.06287}, 2024.

\bibitem{AC24}
O.~Akbar, A. \&~Conlan.
\newblock Towards integrating human-in-the-loop control in proactive intelligent personalised agents.
\newblock In {\em Adjunct Proceedings of the 32nd ACM Conference on User Modeling, Adaptation and Personalization.}, pages 394--398, 2024.

\bibitem{GCC24}
Oihane G{\'o}mez-Carmona, Diego Casado-Mansilla, Diego L{\'o}pez-de Ipi{\~n}a, and Javier Garc{\'\i}a-Zubia.
\newblock Human-in-the-loop machine learning: Reconceptualizing the role of the user in interactive approaches.
\newblock {\em Internet of Things}, 25:101048, 2024.

\bibitem{BCH22}
Alex B{\"a}uerle, {\'A}ngel~Alexander Cabrera, Fred Hohman, Megan Maher, David Koski, Xavier Suau, Titus Barik, and Dominik Moritz.
\newblock Symphony: Composing interactive interfaces for machine learning.
\newblock In {\em Proceedings of the 2022 CHI Conference on Human Factors in Computing Systems}, pages 1--14, 2022.

\bibitem{MHA23}
Eduardo Mosqueira-Rey, Elena Hern{\'a}ndez-Pereira, David Alonso-R{\'\i}os, Jos{\'e} Bobes-Bascar{\'a}n, and {\'A}ngel Fern{\'a}ndez-Leal.
\newblock Human-in-the-loop machine learning: a state of the art.
\newblock {\em Artificial Intelligence Review}, 56(4):3005--3054, 2023.

\bibitem{bengio25}
Yoshua Bengio, Michael Cohen, Damiano Fornasiere, Joumana Ghosn, Pietro Greiner, Matt MacDermott, S{\"o}ren Mindermann, Adam Oberman, Jesse Richardson, Oliver Richardson, et~al.
\newblock Superintelligent agents pose catastrophic risks: Can scientist ai offer a safer path?
\newblock {\em arXiv preprint arXiv:2502.15657}, 2025.

\bibitem{agentic-stat}
{Tom Coshow}.
\newblock Intelligent agents in ai really can work alone. here’s how, 2024.
\newblock Gartner.

\bibitem{AV16}
Sofia Yfantidou, Marios Constantinides, Dimitris Spathis, Athena Vakali, Daniele Quercia, and Fahim Kawsar.
\newblock The state of algorithmic fairness in mobile human-computer interaction.
\newblock In {\em Proceedings of the 25th International Conference on Mobile Human-Computer Interaction}, pages 1--7, 2023.

\end{thebibliography}





\end{document}